\documentclass[%
 reprint,
 amsmath,amssymb,
 aps,
prl,
floatfix,
]{revtex4-1}

\usepackage{graphicx}
\usepackage{subfigure}
\usepackage{dcolumn}
\usepackage{bm}
\usepackage[percent]{overpic}

\begin{document}

\preprint{APS/123-QED}

\title{Vector Magnetometry Exploiting Phase-Geometry Effects in a Double-Resonance Alignment Magnetometer}

\author{Stuart J. Ingleby}
 \email{stuart.ingleby@strath.ac.uk}
\author{Carolyn O'Dwyer}
\author{Paul F. Griffin}
\author{Aidan S. Arnold}
\author{Erling Riis}
\affiliation{%
 Department of Physics, SUPA, Strathclyde University, 107 Rottenrow East, Glasgow, UK\\
}%
\date{\today}

\begin{abstract}
Double-resonance optically pumped magnetometers are an attractive instrument for unshielded magnetic field measurements due to their wide dynamic range and high sensitivity. Use of linearly polarised pump light creates alignment in the atomic sample, which evolves in the local static magnetic field, and is driven by a resonant applied field perturbation, modulating the polarisation of transmitted light. We show for the first time that the amplitude and phase of observed first- and second-harmonic components in the transmitted polarisation signal contain sufficient information to measure static magnetic field magnitude and orientation. We describe a laboratory system for experimental measurements of these effects and verify a theoretical derivation of the observed signal. We demonstrate vector field tracking under varying static field orientations and show that the static field magnitude and orientation may be observed simultaneously, with experimentally realised resolution of 1.7~pT and 0.63~mrad in the most sensitive field orientation.
\end{abstract}

\maketitle

\section{Introduction}
Unshielded magnetic field measurements are a key technique in applications ranging from mineral surveying \cite{Nabighian2005} to archaeology \cite{Ben-Yosef2017}, and the development of compact fT-sensitivity magnetometers \cite{Sheng2017} may lead to significant advances in these applications. The measurement of gradients and curvature in an arbitrarily oriented static magnetic field are of critical importance. The practical difficulties associated with developing portable cryogenic systems for SQUID-based magnetometers makes the development of optically pumped atomic magnetometers attractive. Unshielded optically-pumped gradiometers have been demonstrated recently \cite{Bevilacqua2016}, using a double-resonance magnetometry scheme. In this work we demonstrate a technique for measurement of the full magnetic field vector through the observation of geometry-dependent phase variations in the first- and second-harmonic components of the double-resonance signal. 

In a double-resonance magnetometer, the evolution of atomic spins in a static field $\vec{B}_0$ is interrogated by modulation at a frequency $\omega_{\textrm{RF}}$, with resonant response when $\omega_{\textrm{RF}}$ is equal to the atomic Larmor frequency $\omega_L = \gamma | B_0| $, where $\gamma$ is the gyromagnetic ratio for the probed atomic ground state. Modulation may take the form of oscillating pump light amplitude \cite{Pustelny2006} polarisation \cite{Breschi2014} or frequency \cite{Jimenez-Martinez2010}, or a small oscillating applied field $\vec{B}_{\textrm{RF}}$ \cite{Zigdon2010}. For alkali metal vapour magnetometers operating in the geophysical field range $\omega_{\textrm{RF}} \approx \mathcal{O}(2 \pi \cdot\textrm{100 kHz})$, a convenient frequency range for digitization and software signal analysis, making double-resonance magnetometry a useful technique for uncompensated, portable, unshielded magnetometry, combining high dynamic range and high sensitivity. In order to develop techniques for compact sensors of low cost and power consumption, we use a single monochromatic pump-probe laser beam and apply a small magnetic field perturbation to resonantly drive atomic spin precession. The precessing atomic spins modulate the optical activity of the atomic cell and are detected by measurement of the polarisation of transmitted light. 

Double-resonance sensors have been used widely in scalar field measurements for many years \cite{Bell1961,Bloom1962}. Locking $\omega_{\textrm{RF}}$ to $\omega_{L}$ using the dispersive component of the demodulated signal response allows $|B_0|$ to be determined readily. However, this technique requires that the demodulation phase be set \textit{a priori} and yields only information on the magnitude of $\vec{B}_0$. In addition, signal amplitude in double-resonance magnetometry is highly dependent on the orientation of $\vec{B}_0$ relative to $\vec{B}_{\textrm{RF}}$ and the axis of light propagation. Orientations of $\vec{B}_0$ with zero signal amplitude are known as \textit{dead-zones}. We note that measurement schemes for dead-zone reduction or dead-zone free magnetometry have been demonstrated successfully \cite{Ben-Kish2010}. In this paper we demonstrate a sensor configuration and analysis scheme for determination of $\vec{B}_0$ orientation from the measured phases of the signal contributions observed at $\omega_{\textrm{RF}}$ and $2\cdot\omega_{\textrm{RF}}$. We show that the detected signal can be analysed using an atomic alignment model to determine the magnitude and orientation of $\vec{B}_0$, allowing the full field vector to be inferred.

Various other schemes for vector atomic magnetometry have been demonstrated, including zero-field sensors \cite{Pradhan2016,Seltzer2004}, orthogonal probe lasers \cite{AfachFID2015}, orthogonal pump lasers \cite{Patton2014}, measurement of EIT (electromagnetically induced transparency) resonances \cite{2011Cox} and application of significant slowly varying $\vec{B}_0$ perturbations \cite{Fairweather1972,Lenci2014,Vershovskii2006}. The scheme demonstrated here complements these approaches by addressing some of their practical drawbacks. Zero-field techniques are well-suited for shielded measurements, but lack the dynamic range required for portable unshielded measurements, and additionally require full-field compensation. The use of compensation coils, additional light frequencies or beams and additional $\vec{B}_0$ perturbations add significant hardware overheads and power requirements. We also wished to avoid vector magnetometry schemes requiring sequential measurements under varying field conditions, or observation of free induction decay signals, as these methods require longer sampling times and impose stringent upper limits on the achievable sensor bandwidth.

\section{Theory}

\begin{figure}
\includegraphics[width=\linewidth]{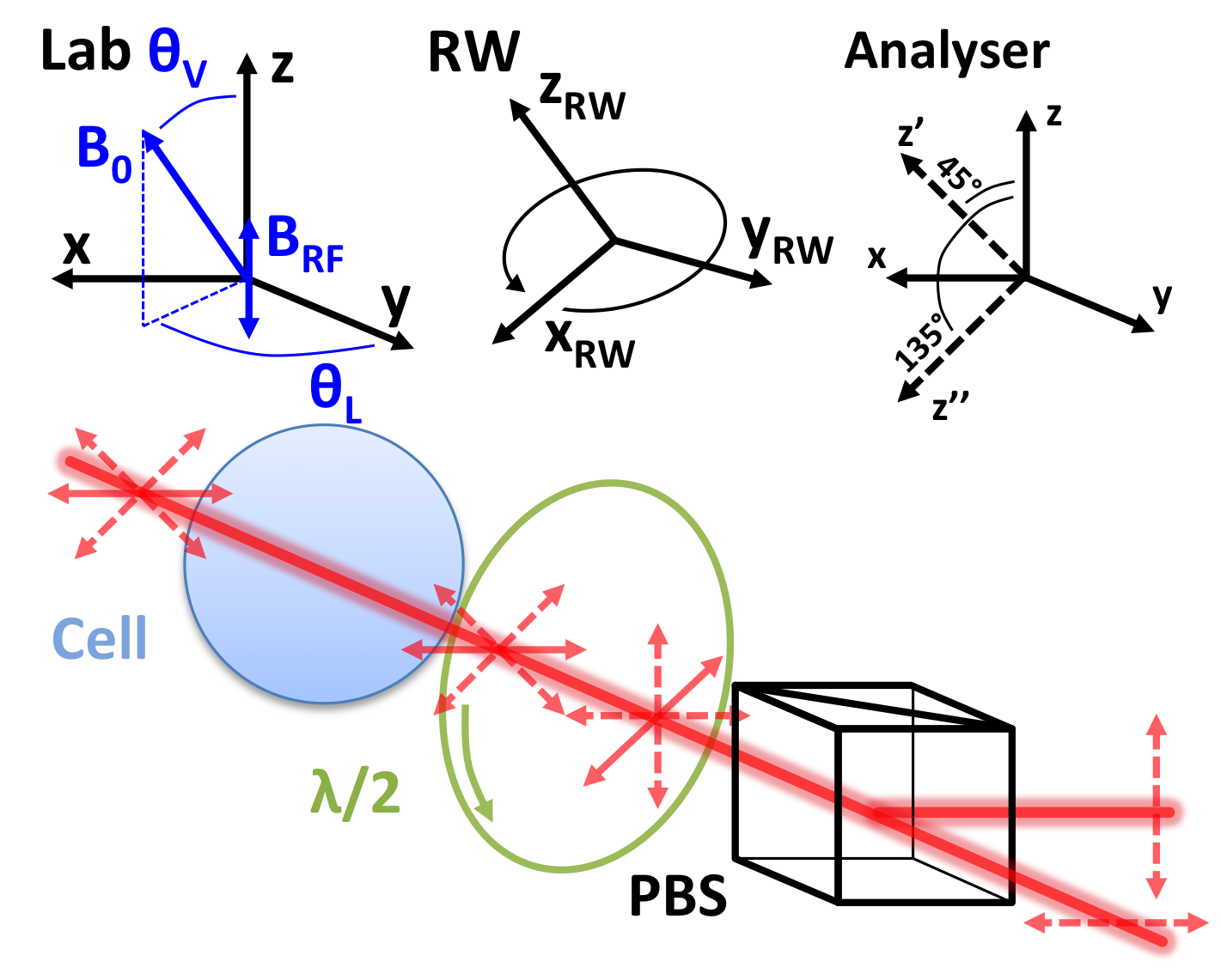}
\caption{Schematic showing the geometry of the optical system and the laboratory (Lab), rotating-wave (RW) and analyser reference frames. The orientation of the static magnetic field $\vec{B}_0$ is described by the spherical polar angles $\theta_V$ and $\theta_L$, and the oscillating magnetic field applied on the $z$-axis. The dashed lines show the linear light polarisation decomposed into orthogonal analysis components, whose intensity difference is measured using a differential photodetector. } 
\label{fig_optics}
\end{figure}

A simple single-beam M\textsubscript{x} magnetometer configuration is used, but the geometry of the static and modulating magnetic fields, atomic sample and analysis optics is critical to the analysis technique and is shown in detail in Figure~\ref{fig_optics}. A half-waveplate is used to balance the detector by rotating the linear polarisation of transmitted light by 45$^{\circ}$, meaning that light which is $x$-polarised at the atoms is equally split by the analyser. The observed differential signal is equal to the difference in transmission of the two orthogonal analysis components separated by the polarising beam splitter. 

The absorption of linear polarisation states by the atomic sample varies with the evolution of polarisation alignment moments in the sample. If the light polarisation axis defines the quantisation axis, then the light absorption coefficient is proportional to 
\begin{equation}
\kappa\propto \frac{A_0}{\sqrt{3}}m_{0,0}-\sqrt{\frac{2}{3}}A_2m_{2,0},
\end{equation}
where the analysing powers $A_0$ and $A_2$ depend on the hyperfine states coupled by the light, and the multipole moments $m_{k,q}$ describe the polarisation of the atomic sample \cite{WeisTheory2006,AuzinshBook2010}.

We can therefore write the observed signal as the difference between the absorption of the two analyser linear polarisation states, as shown in Figure~\ref{fig_optics}. Since the terms in $m_{0,0}$ are invariant under rotations, and cancel in subtraction, the observed differential signal $f(t)$ is proportional to 
\begin{equation}
f(t) = m'_{2,0}(t)-m''_{2,0}(t),
\end{equation}
where $m'$ and $m''$ denote multipole moments describing atomic polarisation alignment in the two orthogonal analysis frames. Rotation \cite{Morrison1987} of these moments into the laboratory frame yields
\begin{equation}
f(t) = \sqrt[]{\tfrac{3}{2}}(m_{2,-1}(t)-m_{2,1}(t)).
\label{eq_ft_signal}
\end{equation}

The dynamic evolution of multipole moments under the static field $\vec{B}_0$ and perturbing field $\vec{B}_{\textrm{RF}}$ can be derived from the Lioville Equation \cite{InglebyOrientation2017}. Steady-state oscillating solutions can be found by setting $\dot{m}_{k,q}=0$ in a frame co-rotating with the perturbing field $\vec{B}_{\textrm{RF}}$ (the rotating wave frame, denoted $m^{\textrm{RW}}_{k,q}$). If the RW frame is chosen such that $\vec{B}_{\textrm{RF}}(t=0)$ is in the $-x$ direction, we can follow the method of \cite{WeisTheory2006}, finding solutions for $m^{\textrm{RW}}_{2,q}$ using
\begin{equation}
\frac{i}{\Gamma}\dot{m}^{\textrm{RW}}_{2,q} = M_{qq'}m^{\textrm{RW}}_{2,q'}+i \bar{m}^{\textrm{RW}}_{2,q},
\end{equation}
where $\Gamma$ is an isotropic spin relaxation rate, $\bar{m}^{\textrm{RW}}_{2,q}$ are moments describing equilibrium magnetisation in the absence of the RF field, and 
\begin{equation}
M_{qq'} = \begin{pmatrix}2x-i&S&0&0&0\\
S&x-i&\sqrt{\tfrac{3}{2}}S&0&0\\
0&\sqrt{\tfrac{3}{2}}S&-i&\sqrt{\tfrac{3}{2}}S&0\\
0&0&\sqrt{\tfrac{3}{2}}S&-x-i&S\\
0&0&0&S&-2x-i\end{pmatrix}.
\end{equation}
For convenience we define the dimensionless quantities $x=(\omega_{\textrm{RF}}-\omega_L)/\Gamma$ and $S=\gamma B^\perp_{\textrm{RF}}/\Gamma$, where $B^\perp_{\textrm{RF}}$ is the component of $\vec{B}_{\textrm{RF}}$ perpendicular to $\vec{B}_0$ and $\gamma$ is the gyromagnetic ratio for the Cs $6^2$S$_{1/2}$~$(F=4)$ ground state.

We have assumed that optical pumping is weak (the optical pumping rate is small compared to the spin relaxation rate $\Gamma$) ensuring that orientation-alignment conversion \cite{Rochester2012} is negligible, atomic spin relaxation is isotropic, and the equilibrium magnetisation $\bar{m}^{\textrm{RW}}_{k,q}$ is aligned with the static field vector $\vec{B}_0$ (\textit{i.e.} $\bar{m}^{\textrm{RW}}_{k,q} = \bar{m}$ for $q=0$, $\bar{m}^{\textrm{RW}}_{k,q} = 0$ otherwise). The magnitude of $\bar{m}$ is proportional to the projection of $m^{\textrm{RW}}_{2,0}$ onto $m^{\textrm{PUMP}}_{2,0}$, where $m^{\textrm{PUMP}}_{k,q}$ are defined in a frame where the quantisation axis is parallel to the polarisation axis of the pump light. 

Steady-state $\dot{m}^{\textrm{RW}}_{2,q} = 0$ solutions for $m^{\textrm{RW}}_{2,q}$ can be found, and so $m_{2,q}(t)$ found by rotation \cite{Morrison1987}. Substitution into (\ref{eq_ft_signal}) yields $f(t)$, with terms in $e^{0\cdot i\omega_{\textrm{RF}} t}$, $e^{1\cdot i\omega_{\textrm{RF}} t}$ and $e^{2\cdot i\omega_{\textrm{RF}} t}$. Similarly to \cite{WeisTheory2006}, we write the amplitude $R$ and phase $\phi$ of the  oscillating responses to $\vec{B}_{\textrm{RF}}$ in the following form; 
\begin{equation}
R(\omega_{\textrm{RF}}) = \frac{A_{1\textrm{f}}\left(x^2(1-2S^2+4x^2)^2+(1+S^2+4x^2)^2\right)^\frac{1}{2}}{(1+S^2+x^2)(1+4S^2+4x^2)}
\label{eq_resonance_fit_x1}
\end{equation}
\begin{equation}
\phi(\omega_{\textrm{RF}}) = \phi^{1\textrm{f}}_0 + \arctan\frac{x(1-2S^2+4x^2)}{1+S^2+4x^2}
\end{equation}
\begin{equation}
R(2\cdot\omega_{\textrm{RF}}) = \frac{A_{2\textrm{f}}\left(9x^2+(1+S^2-2x^2)^2\right)^\frac{1}{2}}{(1+S^2+x^2)(1+4S^2+4x^2)}
\end{equation}
\begin{equation}
\phi(2\cdot\omega_{\textrm{RF}}) = \phi^{2\textrm{f}}_0 + \arctan\frac{3x}{1+S^2-2x^2}.
\label{eq_resonance_fit_y2}
\end{equation}
The on-resonance amplitude $A$ and phase $\phi_0$ of the signal vary with $\theta_L$ and $\theta_V$ as given in Equations~\ref{eq_amplitude_1f}-\ref{eq_phase_2f}.
\begin{equation}
A_{1\textrm{f}}^2 = \bar{m}^2 S^2 \left( (\cos\theta_V \cos\theta_L)^2 + (\cos2\theta_V \sin\theta_L)^2 \right)
\label{eq_amplitude_1f}
\end{equation}
\begin{equation}
A_{2\textrm{f}}^2 = \bar{m}^2 S^4 \left( \left(\tfrac{1}{2}\sin2\theta_V \sin\theta_L\right)^2 + (\sin\theta_V \cos\theta_L)^2 \right)
\label{eq_amplitude_2f}
\end{equation}
\begin{equation}
\tan\phi^{1\textrm{f}}_0 = \frac{-\bar{m}S\cos\theta_V \cos\theta_L}{\bar{m}S\cos2\theta_V \sin\theta_L}
\label{eq_phase_1f}
\end{equation}
\begin{equation}
\tan\phi^{2\textrm{f}}_0 = \frac{2\bar{m}\sin\theta_V \cos\theta_L}{-\bar{m}\sin2\theta_V \sin\theta_L}
\label{eq_phase_2f}
\end{equation}

\section{Test system}

\begin{figure}
\includegraphics[width=\linewidth]{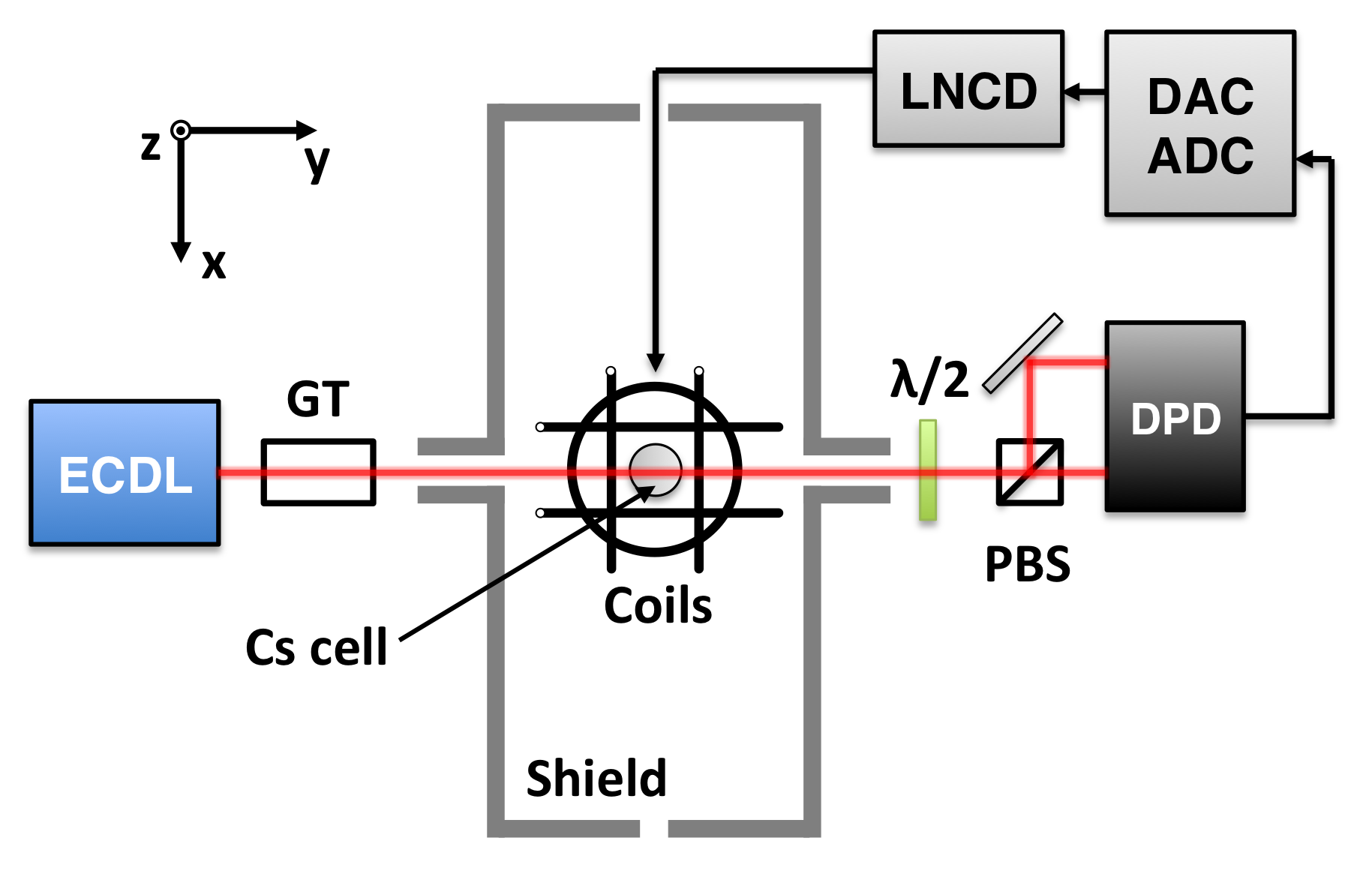}
\caption{Schematic of the experimental system, showing external cavity diode laser (ECDL), Glan-Thompson linear polariser (GT), magnetometer cell, five-layer mu-metal shield, three-axis Helmholtz coils, half-wave plate ($\lambda/2$), polarising beam splitter (PBS), differential photodetector (DPD), low-noise coil driver (LNCD) and data acquisition system (DAC/ADC). The data acquisition system is controlled using a PC (not shown). } 
\label{fig_setup}
\end{figure}

In order to obtain accurate data on the relation of double-resonance signal phase to $\vec{B}_0$ orientation, a shielded test system was used, reducing the effect of background magnetic field noise and allowing fine control of $\vec{B}_0$ orientation. The use of magnetic shielding also allowed us to operate in a low-field regime ($|B_0| \approx 200$~nT), in which the non-linear Zeeman splitting, which leads to systematic shifts in the observed magnetic resonance, is negligible compared to the natural linewidth of the magnetic resonance.

Figure~\ref{fig_setup} shows the test system used, and a detailed hardware description can also be found in \cite{InglebyFieldControl2017}. A spherical room temperature cell of 28~mm diameter containing $^{133}$Cs \cite{CastagnaCells2009} is contained within a five-layer mu-metal shield. Optical access is via a 10~mm diameter axial port and the local static magnetic field at the cell $\vec{B}_0$ is controlled using three pairs of Helmholtz coils driven by six independent software-controlled current supplies. A Helmholtz coil pair on the $z$-axis is used to apply the oscillating perturbation field $\vec{B}_{\textrm{RF}}$. A 1.4 MHz 16-bit DAC/ADC (National Instruments PCIe-6353) is used to generate $\vec{B}_{\textrm{RF}}$ and digitise the differential photodetector signal. Demodulation is carried out in software.

An external-cavity diode laser (New Focus Vortex 6800) provides optical pump/probe light resonant with the $6^2$S$_{1/2}$~$(F=4)$ to $6^2$P$_{1/2}$~$(F=3)$ transition of an external $^{133}$Cs reference cell. This light is linearly polarised along the $x$-axis prior to the magnetometry cell using a Glan-Thompson polariser.

A single magnetic resonance measurement is conducted as follows; following the establishment of the desired $\vec{B}_0$ using the calibrated coil system, an RF modulation signal is generated using the digital-analogue converter. The RF modulation frequency $\omega_{\textrm{RF}}$ is chirped in finite steps. The detector signal response to the modulation signal is synchronously digitised, and a sample segment from each $\omega_{\textrm{RF}}$ step demodulated to obtain the in-phase $X(\omega_{\textrm{RF}})$, $X(2\cdot\omega_{\textrm{RF}})$ and quadrature $Y(\omega_{\textrm{RF}})$, $Y(2\cdot\omega_{\textrm{RF}})$ responses. The sample segments are timed such that each commences in phase with $\vec{B}_{\textrm{RF}}$ and contains an integer number of $\vec{B}_{\textrm{RF}}$ periods. Sample segment length is kept approximately constant for all $\omega_{\textrm{RF}}$, and each sample segment is preceded by a pre-trigger segment of fixed duration, to allow the steady-state oscillating response to $\vec{B}_{\textrm{RF}}(\omega_{\textrm{RF}})$ to be measured. 

\begin{figure}
\begin{tabular}{cc}
\includegraphics[width=0.5\linewidth]{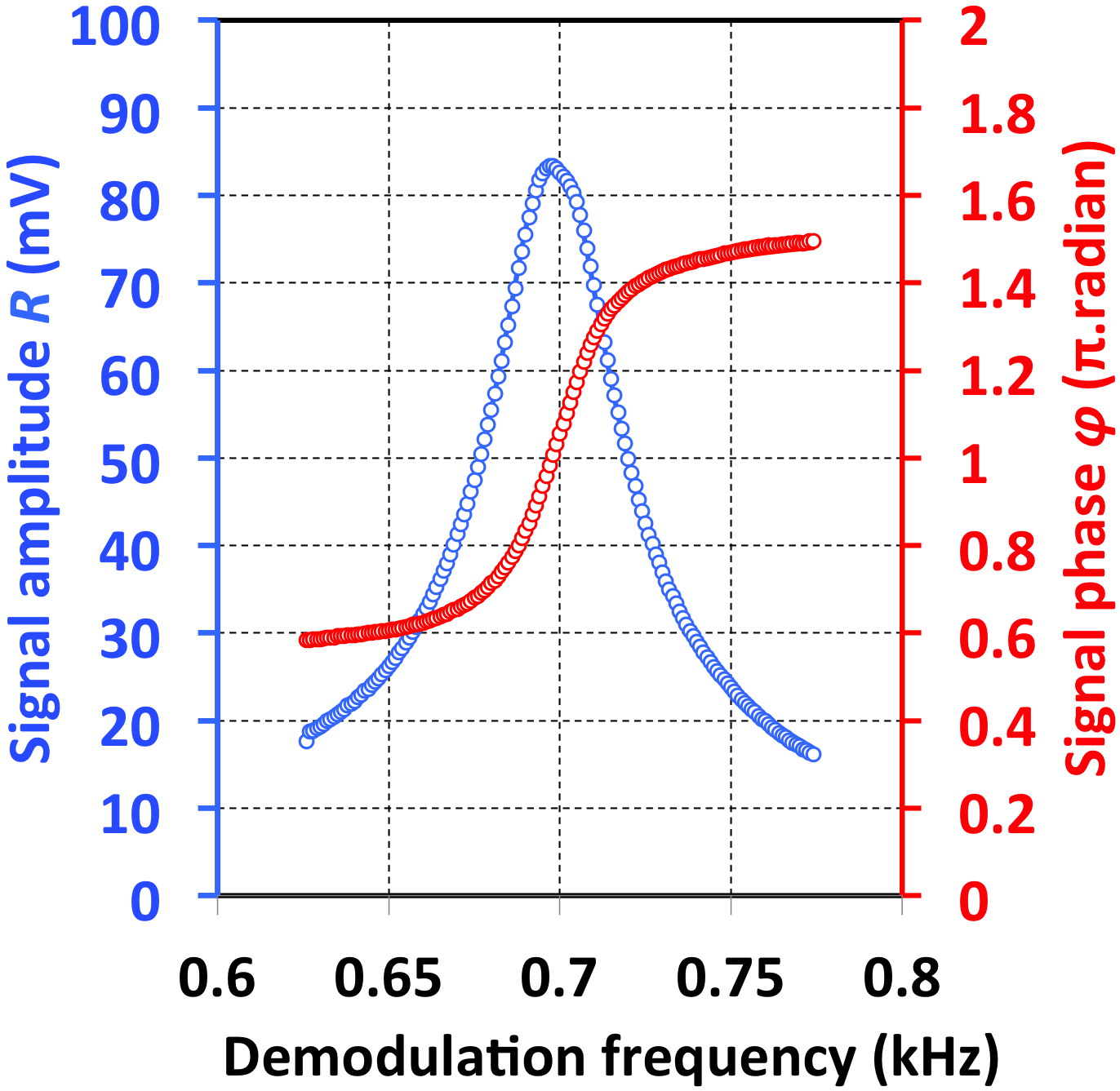} &
\includegraphics[width=0.5\linewidth]{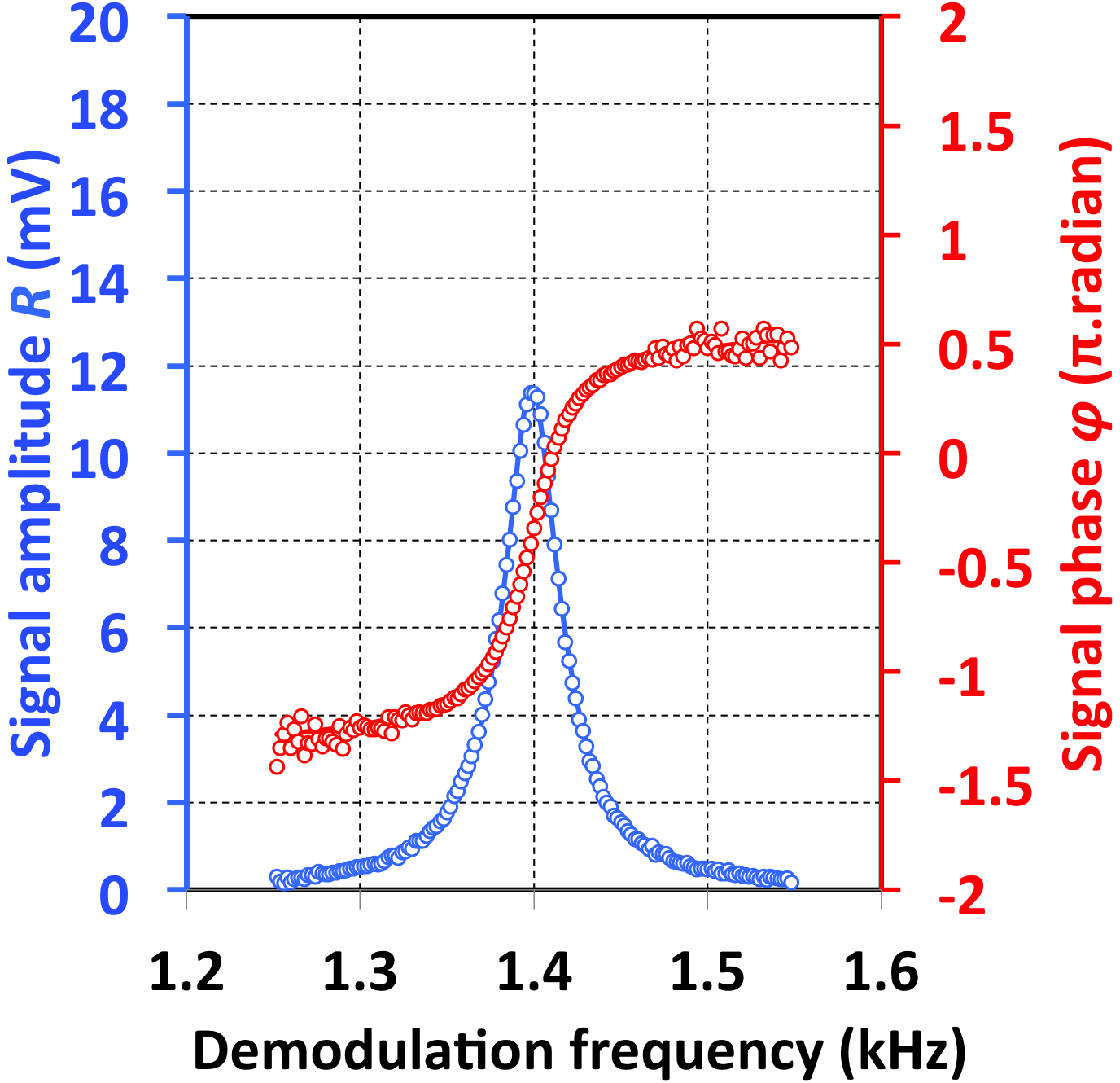}
\end{tabular}
\caption{A measured and fitted magnetic resonance, taken with $|B_0|=$~200~nT applied at $\theta_V = 118^{\circ}$ , $\theta_L = 101^{\circ}$ and $|B_{\textrm{RF}}| =$~1.5~nT. A total of 150 segments of data are taken, with segment sample time 20~ms. \textit{Left:} amplitude ($R$) and phase ($\phi$) components of the first-harmonic demodulated signal. \textit{Right:} amplitude ($R$) and phase ($\phi$) of the second-harmonic demodulated signal. The data are fitted with Equations~\ref{eq_resonance_fit_x1}~-~\ref{eq_resonance_fit_y2}, yielding $\omega_L=2\pi\cdot$~699.60(2)~Hz, $\Gamma=$~12.1(1)~Hz, $A_{1\textrm{f}}=$~107.4(7)~mV, $\phi^{1\textrm{f}}_0=$~0.9672(9)~$\pi$.rad, $\phi^{2\textrm{f}}_0=-0.383(6)\pi$.rad and $\Omega=$~2.89(8)~Hz.} 
\label{fig_resonance}
\end{figure}

Figure~\ref{fig_resonance} shows measured signal amplitude $R\equiv \sqrt[]{X^2+Y^2}$ and phase $\phi \equiv \arctan(X/Y)$ for data demodulated at $\omega_{\textrm{RF}}$ and $2\cdot\omega_{\textrm{RF}}$. Least-squares fits of Equations~\ref{eq_resonance_fit_x1} -~\ref{eq_resonance_fit_y2} (these resonance shapes and the underlying physical model are described in detail below) are used to estimate the Larmor frequency $\omega_L$, spin relaxation rate $\Gamma$, on-resonance signal amplitude $A$ and phase $\phi_0$, and magnetic Rabi rate $\Omega$.

\section{Static Field Calibration}

To achieve precise control of $\vec{B}_0$, allowing measurement of orientational effects, the static field generating coils are calibrated by measurement of the Larmor frequency under varying orientations of the applied field. The method for initial coil calibration is described in \cite{InglebyFieldControl2017}. For a given application of the applied field $\vec{B}^{\textrm{APP}}$, the magnitude of the measured field $|B^{\textrm{MEAS}}|=\omega_L / \gamma$ is determined by fitting Equations~\ref{eq_resonance_fit_x1} - \ref{eq_resonance_fit_y2} to the demodulated data $R(\omega_{\textrm{RF}})$ and $\phi(\omega_{\textrm{RF}})$.

Following the initial calibration, fine coil calibration is carried out by orienting $\vec{B}^\textrm{APP}$ in 1646 orientations, spaced with equal angular coverage over the full solid angle, and performing a weighted fit to the observed distribution of $|B^{\textrm{MEAS}}|$ with
\begin{equation}
|B^{\textrm{MEAS}}| = \sqrt[]{\sum_i (\epsilon_i + a_i B^\textrm{APP}_i)^2},
\end{equation}
where $\epsilon$ is the background field and $a$ is a dimensionless coil calibration factor. The calibration and offset of each coil can then be corrected by the best-fit parameters $a_i$ and $\epsilon_i$. The uncertainties in the fit $\delta a$ and $\delta \epsilon$ can be used to estimate the tolerances in the magnitude and orientation of $\vec{B}_0$, $\delta B_0$ and $\delta\theta$, by assuming that the total field uncertainty, estimated by $\delta B_0 = |\delta \epsilon| + |B_0||\delta a|$, is perpendicular to $\vec{B}_0$, yielding $\delta\theta \approx \delta B_0 / |B_0|$ for $\delta B_0 \ll |B_0|$. 

\begin{figure}
\begin{tabular}{c}
\includegraphics[width=\linewidth]{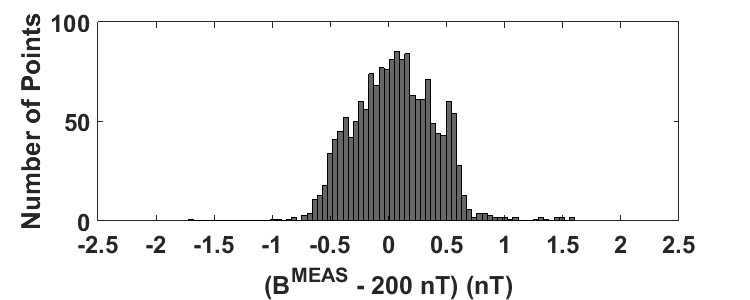} \\
\includegraphics[width=\linewidth]{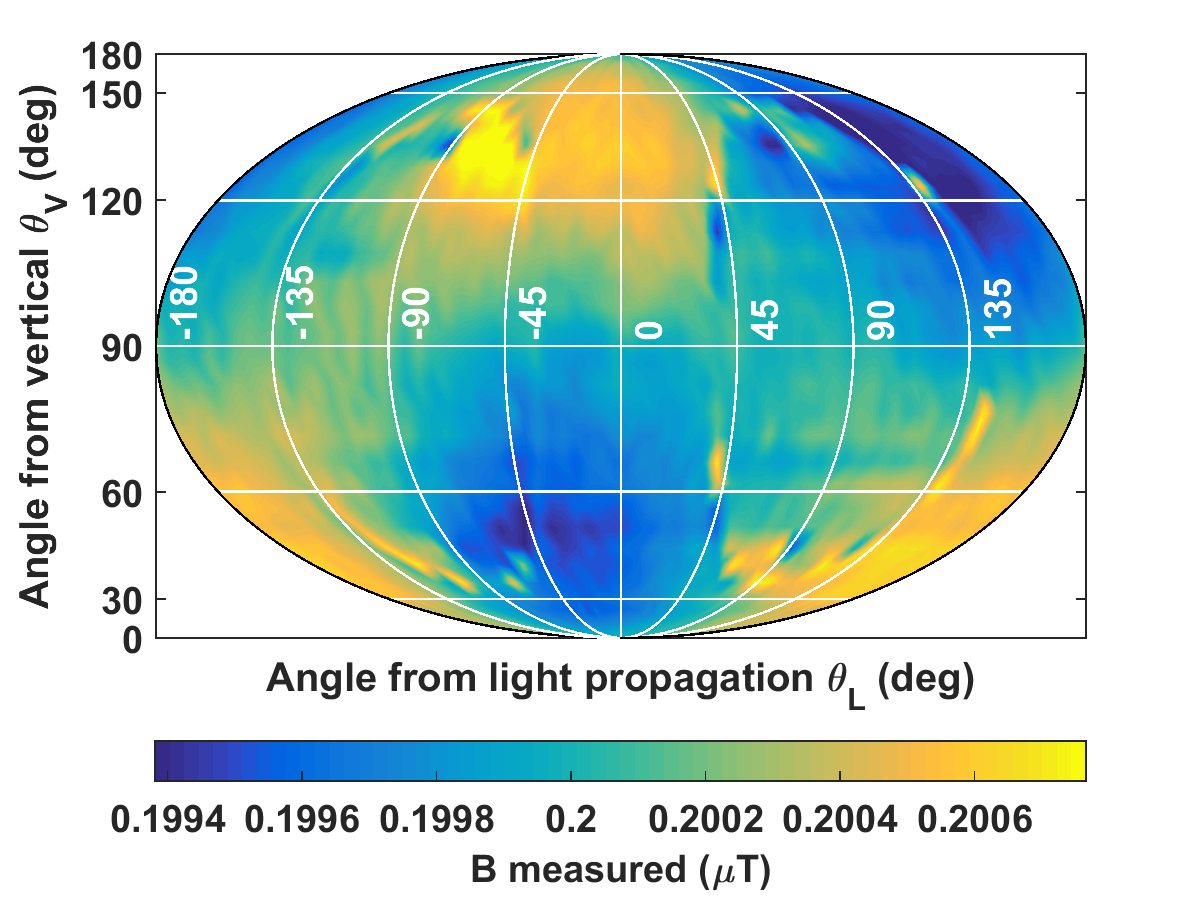}
\end{tabular}
\caption{Measured magnitude of $\vec{B}_0$, determined from measurements of $\omega_L$ at 1646 different $\vec{B}_0$ orientations, evenly covering the full solid angle. \textit{Top:} distribution of measured $|B_0|$ around desired field magnitude of 200~nT. The RMS spread of $|B_0|$ is 302~pT. \textit{Bottom:} angular distribution of $|B_0|$.} 
\label{data_homogeneity}
\end{figure}

Table~\ref{tab_calibration} gives the calibration parameter uncertainties for the final coil calibration, and Figure~\ref{data_homogeneity} shows the measured value of $|B_0|$ over the full solid angle for the subsequent field vector measurements. In order to render heading-error effects due to non-linear Zeeman splitting negligible, a field magnitude of $|B_0| \approx 200$~nT is used throughout. From the calibration uncertainties we estimate tolerances of $\delta |B_0| = 54$~pT and $\delta \theta = 0.27$~mrad. The RMS spread of observed magnitudes $\delta |B_0|^{RMS}$ from Figure~\ref{data_homogeneity} is 302~pT. Although the difference between $\delta |B_0|$ and $\delta |B_0|^{RMS}$ is indicative of some remaining non-normal (\textit{i.e.} anisotropic, systematic) contributions to $\vec{B}_0$ discrepancies, we can still be confident that $\vec{B}_0$ can be set with orientational fidelity in the mrad range.

\begingroup
\begin{table}
\caption{Uncertainties in coil calibration parameters from final calibration fit for $|B_0| = 200$~nT. The resulting tolerances in $\vec{B}_0$ magnitude and orientation are $\delta |B_0| = 54$~pT and $\delta \theta = 0.27$~mrad.}
\begin{tabular}{ c | c c } 
Coil axis & $\delta a$ ($10^{-5}$) & $\delta \epsilon$ (pT)\\
 \hline
 x & 7.2 & 11 \\ 
 y & 10.0 & 14 \\ 
 z & 9.7 & 14 \\
\end{tabular}
\label{tab_calibration}
\end{table}
\endgroup

\section{Vector Field Measurements}

\begin{figure}[t]
\begin{overpic}[width=0.5\textwidth]{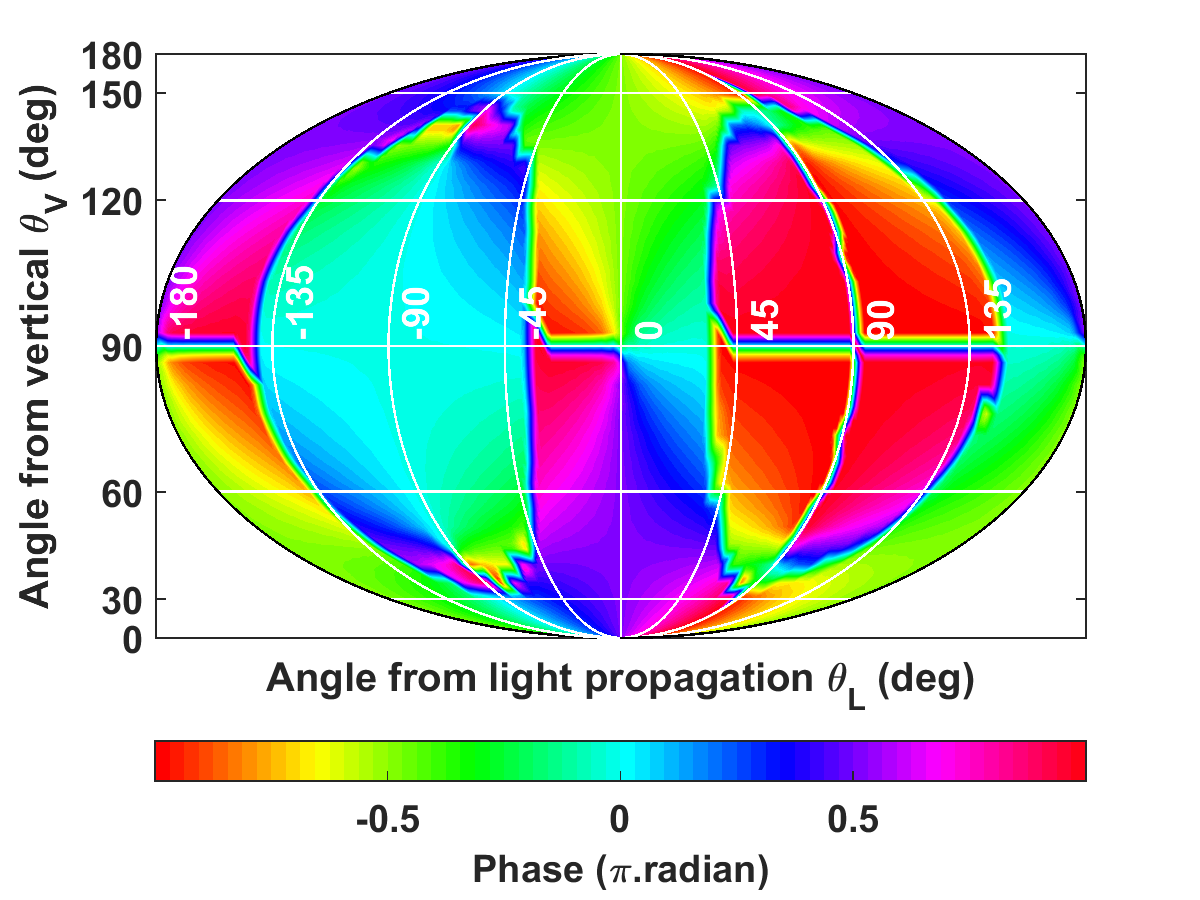}
\put(65,53){\includegraphics[scale=0.2]{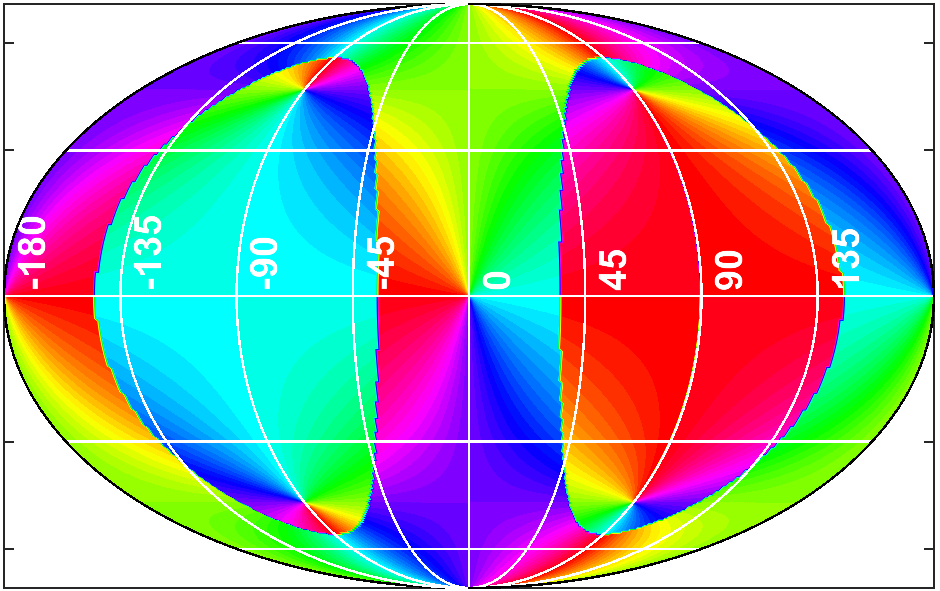}}
\end{overpic}
\begin{overpic}[width=0.5\textwidth]{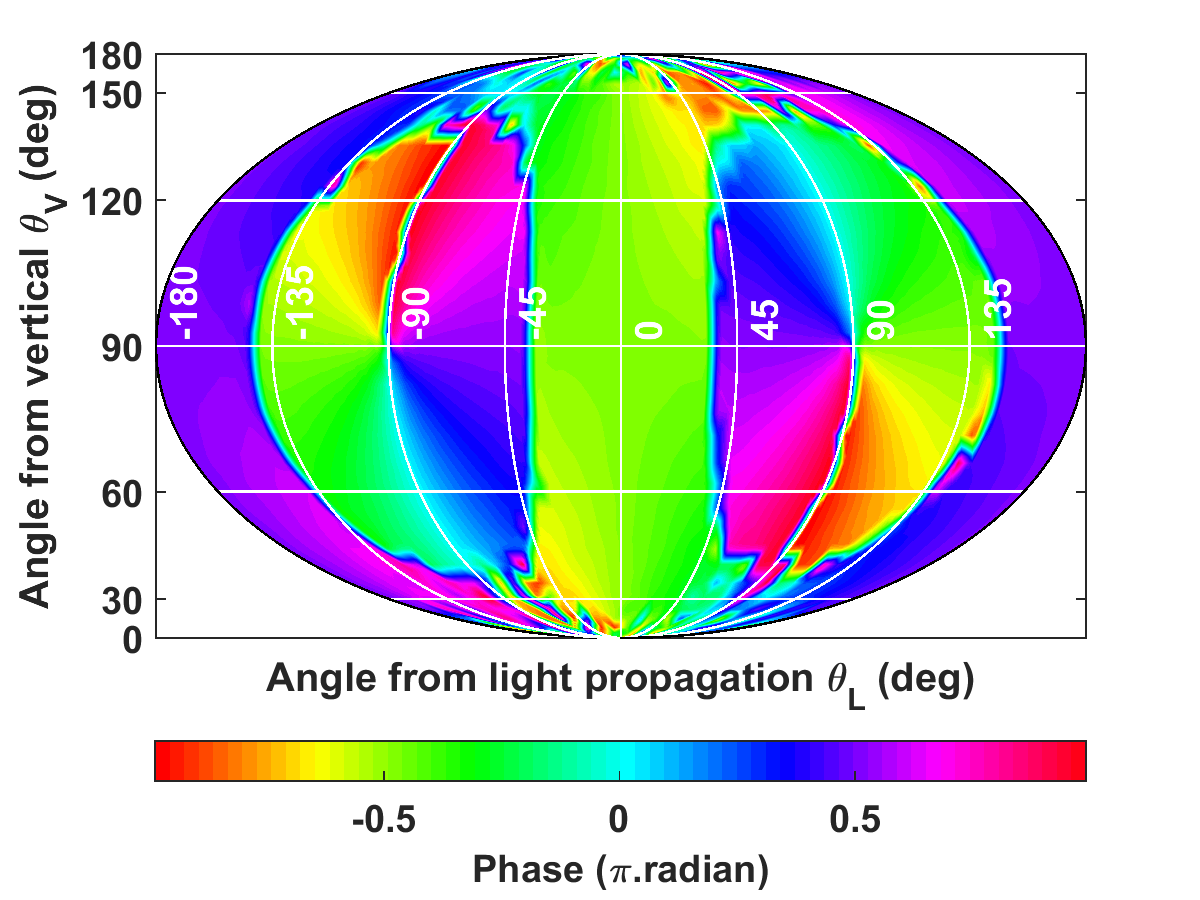}
\put(65,53){\includegraphics[scale=0.2]{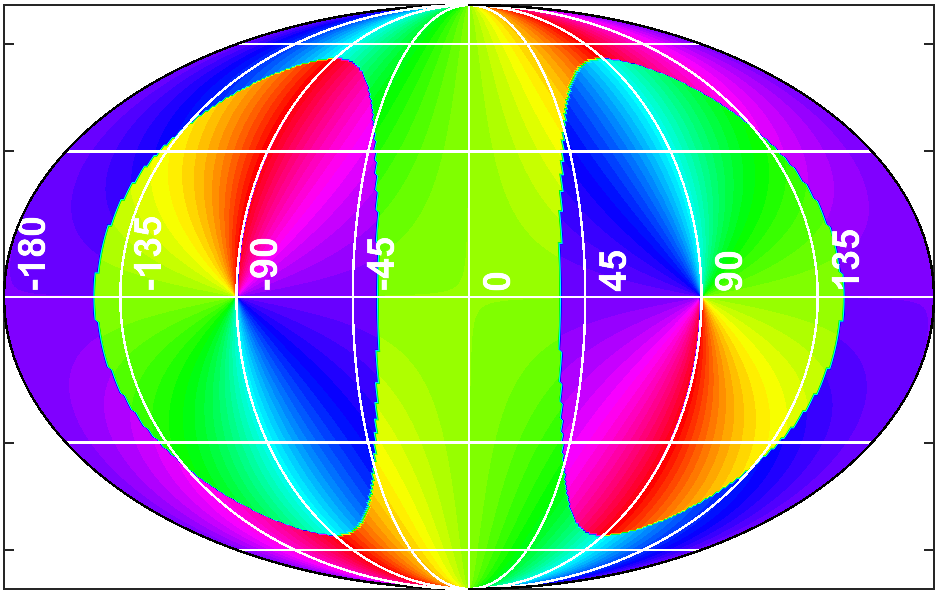}}
\end{overpic}
\caption{Observed and (\textit{inset}) calculated distributions of on-resonance signal phase with variation of $\vec{B}_0$ orientation over full solid angle. \textit{Top}: first-harmonic phase $\phi^{1\textrm{f}}_0$. \textit{Bottom}: second-harmonic phase $\phi^{2\textrm{f}}_0$. Calculated distributions are found using Equations~\ref{eq_phase_1f}~-~\ref{eq_phase_2f}.} 
\label{data_phase}
\end{figure}

\begin{figure}[t]
\begin{overpic}[width=0.5\textwidth]{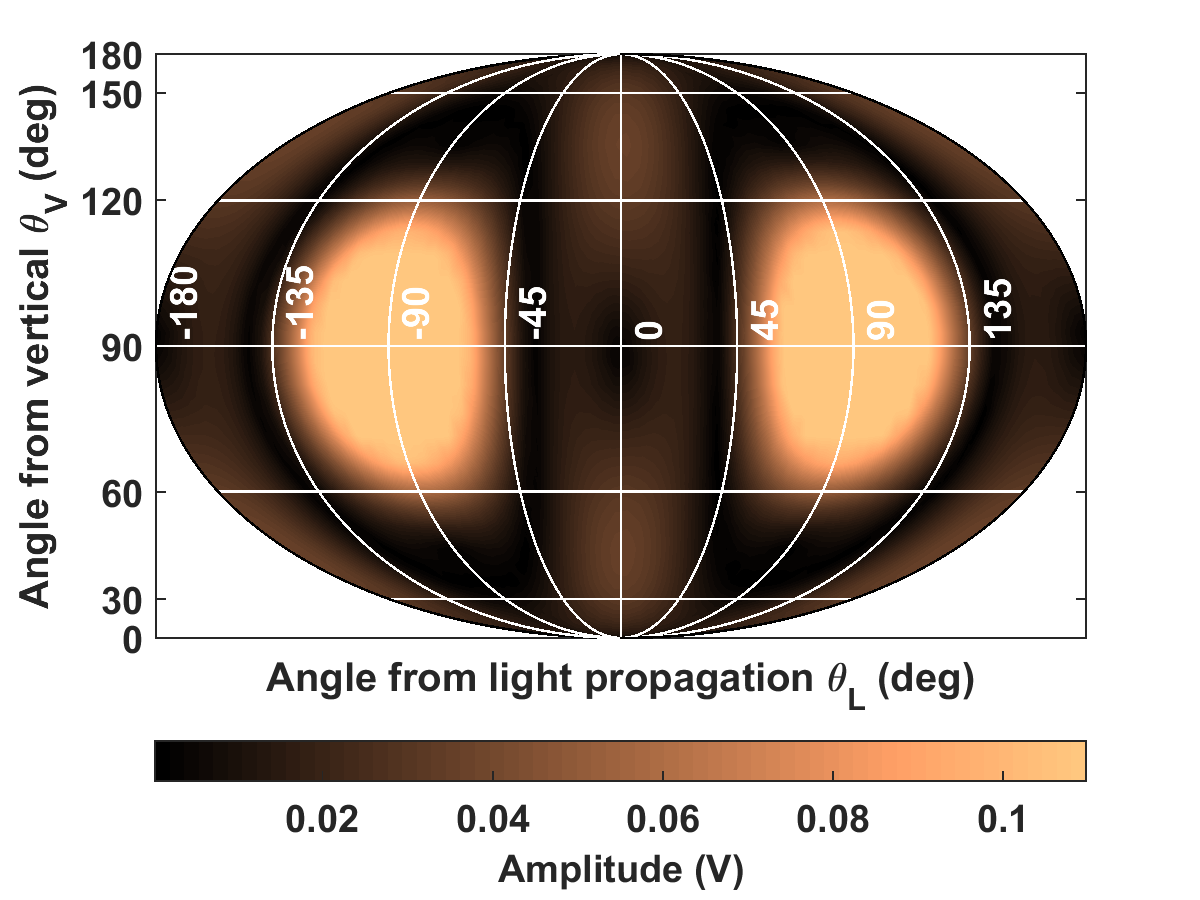}
\put(65,53){\includegraphics[scale=0.2]{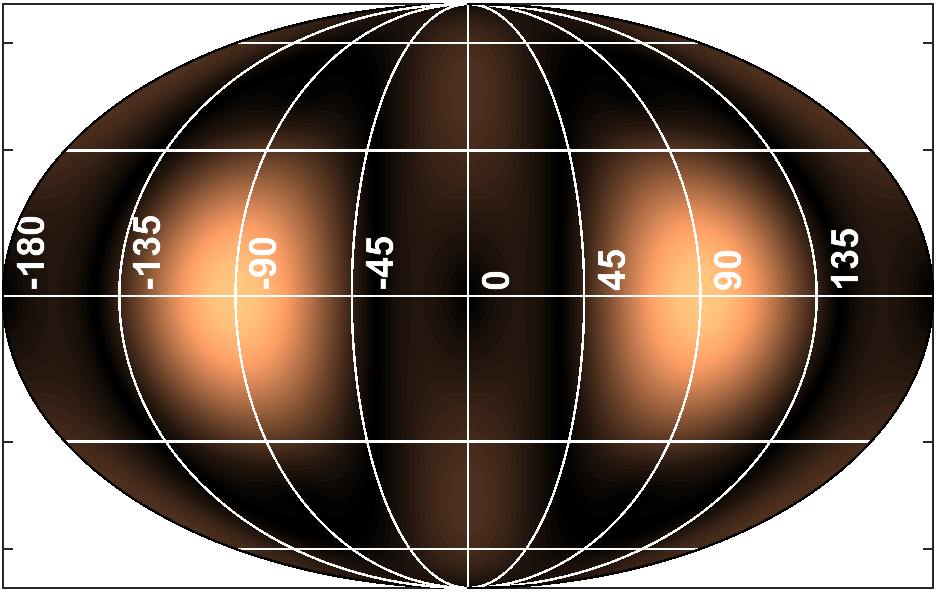}}
\end{overpic}
\begin{overpic}[width=0.5\textwidth]{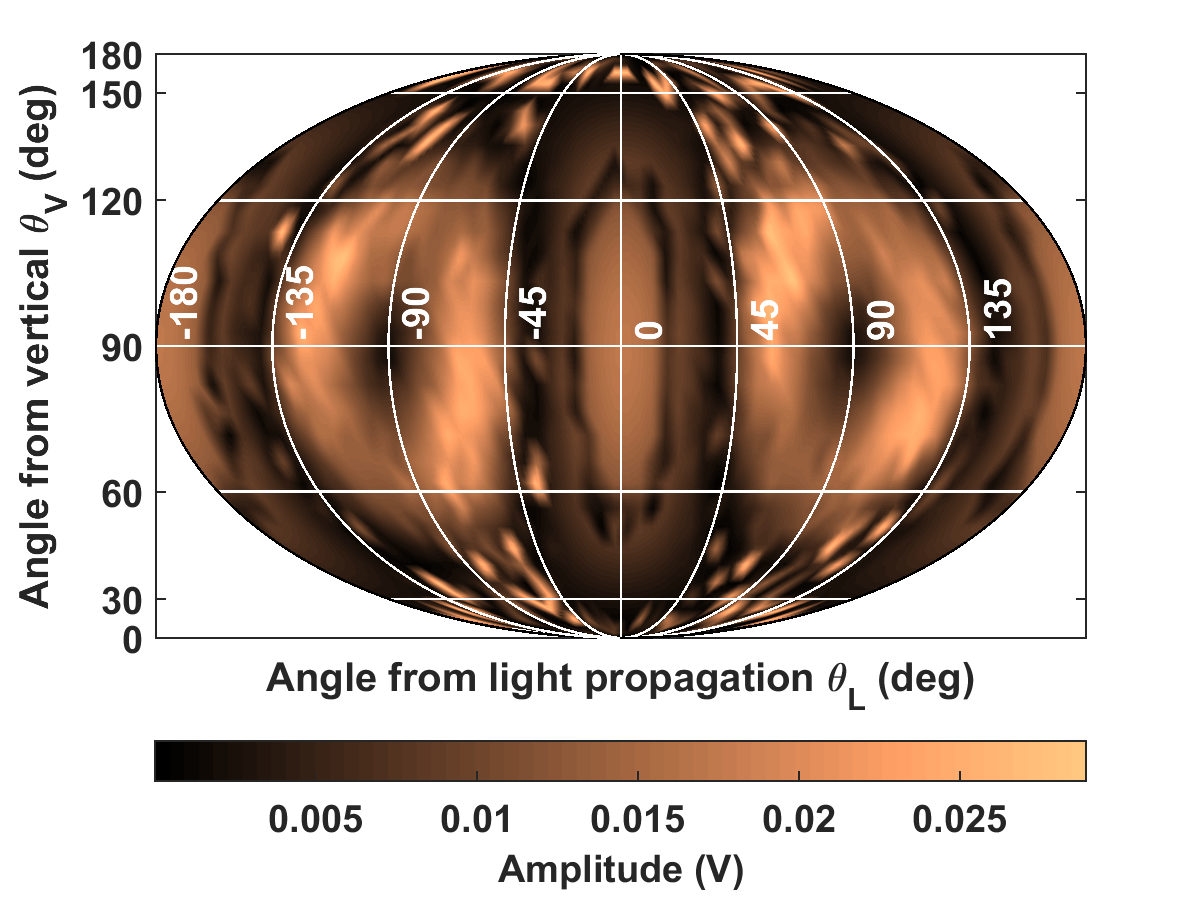}
\put(65,53){\includegraphics[scale=0.2]{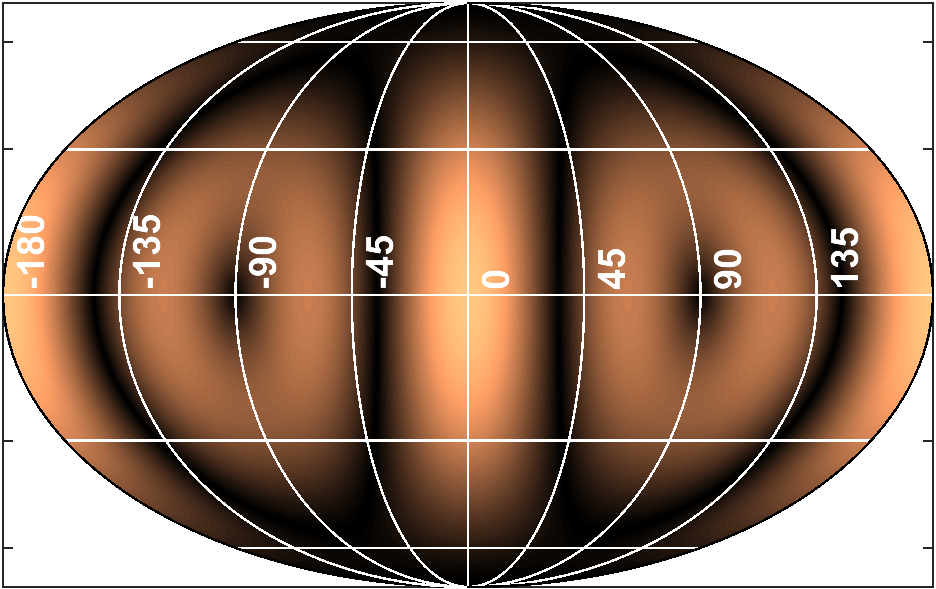}}
\end{overpic}
\caption{Observed and (\textit{inset}) calculated distributions of on-resonance signal amplitude with variation of $\vec{B}_0$ orientation over full solid angle. \textit{Top}: first-harmonic amplitude $A_{1\textrm{f}}$. \textit{Bottom}: second-harmonic amplitude $A_{2\textrm{f}}$. Calculated distributions are found using Equations~\ref{eq_amplitude_1f}~-~\ref{eq_amplitude_2f}.} 
\label{data_amplitude}
\end{figure}

Equations~\ref{eq_amplitude_1f}~-~\ref{eq_phase_2f} indicate a strong dependency between the on-resonance signal components and the orientation of $\vec{B}_0$. Using the calibrated field control described above, automated scans of $\vec{B}_0$ were carried out. Each scan consists of 1646 orientations of $\vec{B}_0$, spread over the full solid angle with approximately even angular distribution. At each $\vec{B}_0$ orientation, a magnetic resonance measurement was carried out, and a fit to the resulting data using Equations~\ref{eq_resonance_fit_x1}~-~\ref{eq_resonance_fit_y2} used to obtain best-fit values and uncertainties for $\phi^{1\textrm{f}}_0$, $\phi^{2\textrm{f}}_0$, $A_{1\textrm{f}}$ and $A_{2\textrm{f}}$. The results of this measurement are shown in Figures~\ref{data_phase}~-~\ref{data_amplitude}. Good agreement was found between the on-resonance signal components in the measured data and Equations~\ref{eq_amplitude_1f}~-~\ref{eq_phase_2f}. 

\begin{figure*}
\begin{overpic}[width=\textwidth]{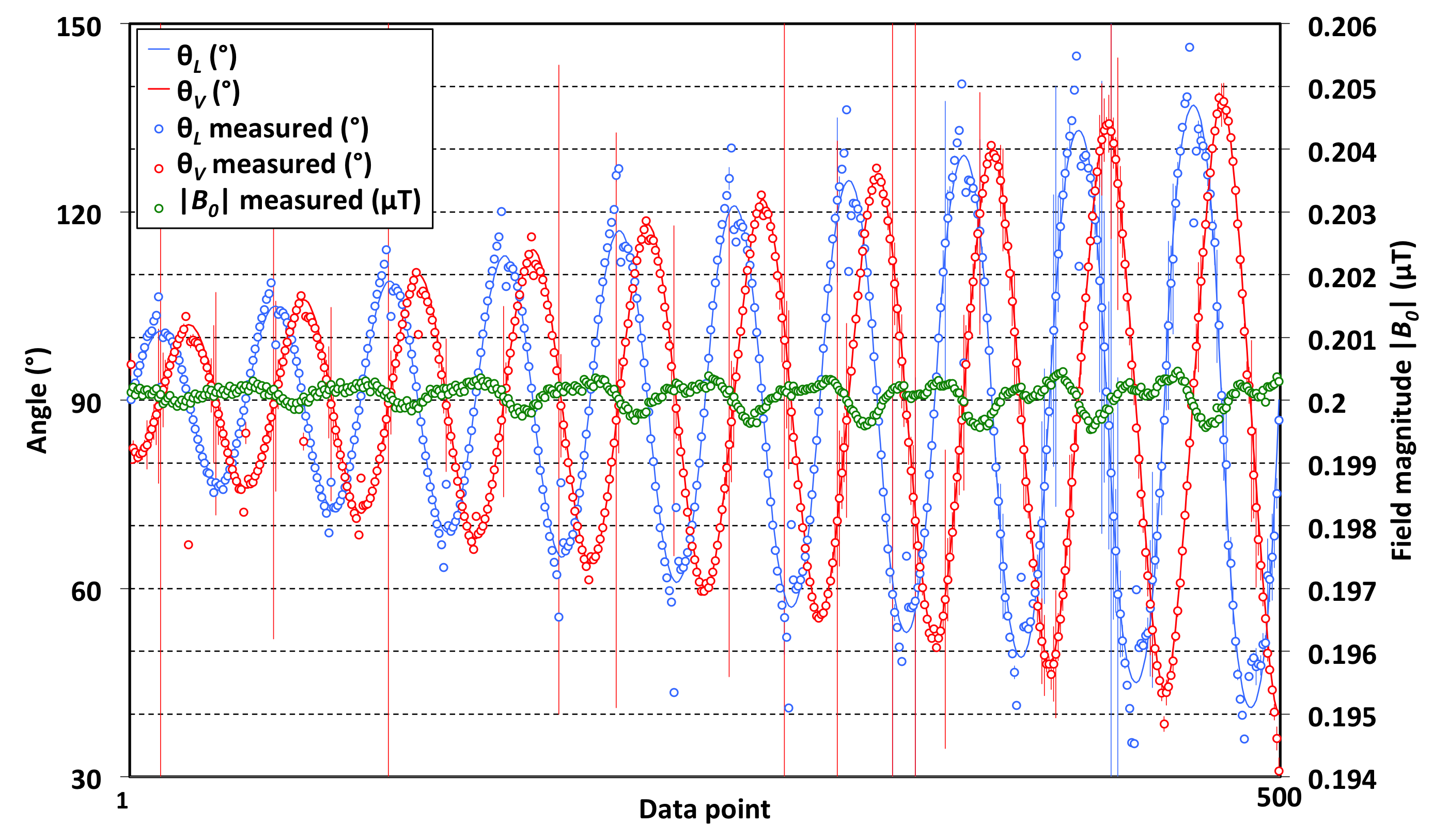}
\put(9.24,4.5){\includegraphics[scale=0.25]{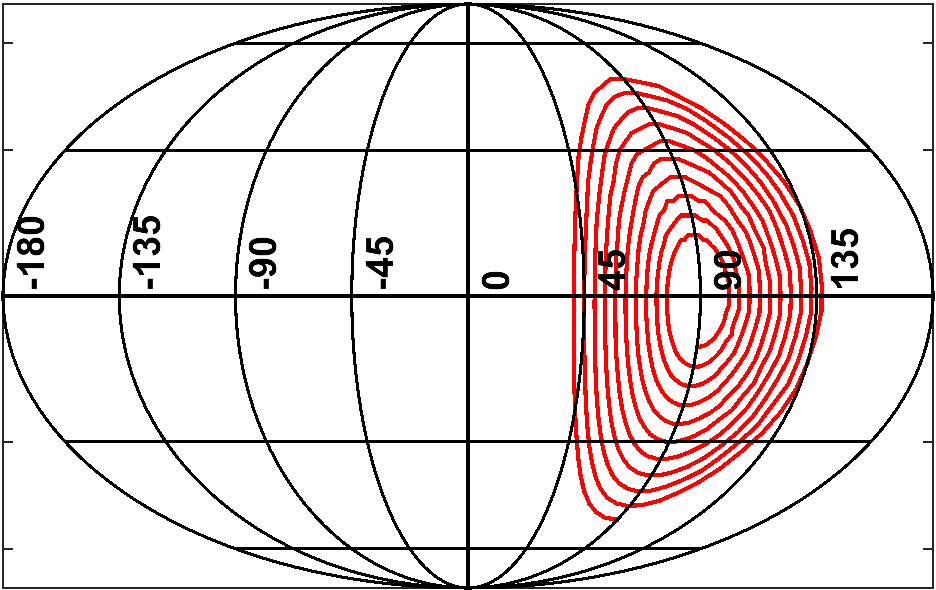}}
\end{overpic}
\caption{Measured and set magnetic field magnitude and orientation for a wide range of $\vec{B}_0$ orientations. Observed values and uncertainties for $\theta_V$ and $\theta_L$ are calculated from $\phi^{1\textrm{f}}_0$ and $\phi^{2\textrm{f}}_0$ using Equations~\ref{eq_theta_V} and~\ref{eq_theta_L}. The point of best angular resolution is measured at ($|B_0| = 199.8445(17)$~nT , $\theta_V = 100.986(27)^{\circ}$ , $\theta_L = 118.198(24)^{\circ}$ ). The inset in the lower left corner shows the contour described by successive $\vec{B}^{\textrm{APP}}$ orientations, plotted using the same projection as Figures~\ref{data_phase} and \ref{data_amplitude}.} 
\label{data_contour_scan}
\end{figure*}

We note the dependence of the first- and second-harmonic on-resonance signal phases $\phi^{1\textrm{f}}_0$ and $\phi^{2\textrm{f}}_0$ on the orientation of $\vec{B}_0$. From Equations~\ref{eq_phase_1f} - \ref{eq_phase_2f} we can derive Equations~\ref{eq_theta_V} - \ref{eq_theta_L} for $\theta_V$ and $\theta_L$. 
\begin{equation}
\tan^2 \theta_V = 1 - \frac{\tan\phi^{2\textrm{f}}_0}{\tan\phi^{1\textrm{f}}_0}
\label{eq_theta_V}
\end{equation}
\begin{equation}
\tan \theta_L = \frac{-\cos\theta_V}{\cos2\theta_V\tan\phi^{1\textrm{f}}_0}
\label{eq_theta_L}
\end{equation}
By measuring the resonant response of the detector signal, demodulating to obtain the first- and second-harmonic signal amplitude and phase, and fitting to determine the Larmor frequency and on-resonance phases, we can calculate $\theta_V$ and $\theta_L$ and make a full-vector measurement of $\vec{B}_0$. 

Figure~\ref{data_contour_scan} shows calculated $\theta_V$ and $\theta_L$ for a range of $\vec{B}_0$ orientations defined using the calibrated Helmholtz coil system. The range of orientations is shown as an inset to Figure~\ref{data_contour_scan} and was chosen to scan over the zone of high signal amplitude around the light polarisation (\textit{x}-) axis. At each point the first- and second-harmonic resonance responses are measured for a range of $x$ and fitted using Equations~\ref{eq_resonance_fit_x1} - \ref{eq_resonance_fit_y2}. The on-resonance phases $\phi^{1\textrm{f}}_0$ and $\phi^{2\textrm{f}}_0$ are free parameters in this model, and the fit uncertainties are propagated through Equations~\ref{eq_theta_V} - \ref{eq_theta_L} to give the uncertainties in $\theta_V$ and $\theta_L$. The point of highest observed angular resolution has uncertainties of $\delta|B_0| = 1.7$~pT, $\delta\theta_V = 0.027^{\circ}$, $\delta\theta_L = 0.024^{\circ}$, giving an overall angular resolution at this point of $\delta\theta = 0.036^{\circ}$~(0.63 mrad).

\section{Conclusions}
The measurement of complementary field orientation information using a hitherto-scalar double-resonance magnetometry technique has clear potential for impact in practical measurements of arbitrarily oriented fields. Existing three-axis magnetometer data is often transformed to derive data on field magnitude, declination and inclination. In this work we demonstrate a scheme for independent measurement of the field vector in this spherical polar basis, while also exploiting the precise and accurate measurement of field magnitude possible with the double-resonance technique. The single-beam, RF-modulated detection scheme used is imminently suitable for scalable, portable devices.

The data shown in Figure~\ref{data_contour_scan} demonstrate resolution of the magnetic field magnitude at the pT-level and magnetic field orientation at the sub-mrad level. The variation of the measured field magnitude and orientation from the expected field magnitude (200~nT) and orientation (solid lines) exposes residual calibration errors in the Helmholtz coil system, which can set $\vec{B}_0$ with tolerances in the 100-pT and few-mrad ranges. The general validity of the phase-orientation effects derived from theory and observed in Figure~\ref{data_phase} are not called into question, but a more stringent test of the absolute accuracy of the field orientation measurement will require improvements to the hardware of the Helmholtz coil system, including improved design tolerances on the coil geometry (currently at the 100-micron level), improved linearity of the coil current drivers and associated DACs and increased detector signal-to-noise, which would also improve the resolution of both the calibration and vector field data. 

The double-resonance scheme presented also has some drawbacks in the implementation of practical sensors, which may form the context for further work. We observe dead-zones, both where signal amplitude falls to zero (dark regions in Figure~\ref{data_amplitude}) and angular dead-zones; orientations for which the observed signal phase has no variation with field orientation $\partial\phi_0/\partial\theta = 0$. These angular dead-zones do not necessarily coincide with the signal-amplitude dead-zones, and can be seen in Figure~\ref{data_contour_scan} as angular data points with very high uncertainties. A further drawback of this technique is the requirement that magnetic detuning $x$ be measured independently from phases-on-resonance $\phi^{\textrm{1f}}_0$ and $\phi^{\textrm{2f}}_0$. In this work we met this requirement at the expense of bandwidth by measuring and fitting a $\omega_{\textrm{RF}}$ frequency sweep at each data point.

To conclude, we have demonstrated a new analysis technique for double-resonance alignment magnetometry that can be used to implement vector magnetometry using a scalar device. No additional lasers or field-generating coils are required, and the vector field sensitivity achieved using this technique could be further enhanced by rapid independent measurement of $x$, $\phi^{\textrm{1f}}_0$ and $\phi^{\textrm{2f}}_0$, allowing the field vector $\vec{B}_0 (t)$ to be determined with high bandwidth.
\section{Acknowledgements}
The authors would like to thank Prof. Antoine Weis and Dr. Victor Lebedev of Fribourg University for supplying the Cs vapour cell used in this work. This work was funded by the UK Quantum Technology Hub in Sensing and Metrology, EPSRC (EP/M013294/1). The data shown in this paper is available for download at  
http://dx.doi.org/10.15129/5f63cec2-e674-4e42-b923-7550e28d860f.

\bibliography{main.bib}
\end{document}